\documentclass[pra,twocolumn,amsmath,amssymb]{revtex4-2}
\usepackage{physics}
\usepackage{bm}
\usepackage{dcolumn}
\usepackage{graphicx}
\usepackage{xcolor}
\usepackage[colorlinks,urlcolor=blue,citecolor=blue,linkcolor=blue]{hyperref}

\begin{document}

\title{Environment-modified three-body energy transfer}
\author{Madeline C.~Waller and Robert Bennett}
\affiliation{School of Physics \& Astronomy, University of Glasgow, Glasgow G12 8QQ, United Kingdom}
\date{\today}
\begin{abstract}
  Resonant energy transfer from a donor to an acceptor is one of the most basic interactions between atomic and molecular systems. In real-life situations, the donor and acceptor are not isolated but in fact coupled to their environment and to other atoms and molecules. The presence of a third body can modify the rate of energy transfer between donor and acceptor in distinctive and intricate ways, especially when the three-site system is itself interacting with a larger macroscopic background such as a solvent. The rate can be calculated perturbatively, which ordinarily requires the summation of very large numbers of Feynman-like diagrams. Here we demonstrate a method based on canonical perturbation theory that allows us to reduce the computational effort required, and use this technique to derive a formula for the rate of three-body resonance energy transfer in a background environment. As a proof-of-principle, we apply this to the situation of a dimer positioned near a dielectric interface, with a distant third molecule controlling the rate, finding both enhancement or suppression of the rate depending on system parameters.
\end{abstract}

\maketitle

\section{Introduction}
The transport of energy between atoms and molecules is important in many diverse areas of science. It is present as a fundamental process in the transport of energy in plants, and has applications in, for example, artificial light-harvesting \cite{mohapatra_forster_2018} and the “spectroscopic ruler”, a technique used to estimate intermolecular distances within macromolecules \cite{stryer_energy_1967}. A closely related process, Interatomic Coulombic Decay (ICD) \cite{cederbaum_giant_1997}, could also be relevant in radiation biology \cite{boudaiffa_resonant_2000}. The mechanism that governs the transfer of energy from one atom/molecule (donor) to another (acceptor) depends chiefly on the strength of the light-matter coupling and the interatomic/intermolecular distance. When the distances are ultra-short, the resulting energy transfer is governed by Dexter theory \cite{dexter_theory_1953}, where electronic orbitals overlap allowing the electrons to migrate between molecules. For longer distances where the electronic orbitals no longer overlap, energy transfer is instead mediated by a photon, with electrons of the donor and acceptor remaining bound to the nuclei. For near field interactions where light-matter coupling is weak relative to the intramolecular (e.g. vibrational) coupling, this energy transfer is governed by Förster theory \cite{forsterZwischenmolekulareEnergiewanderungUnd1948}, leading to the well-known $R^{-6}$ dependence on the separation distance $R$. For far field interactions, relativistic properties of the mediating photon become significant, leading to an $R^{-2}$ distance dependence \cite{power_new_1993}. When intramolecular couplings are comparable to (or weaker than) intermolecular ones, completely different approaches must be taken (see, e.g., \cite{yang_influence_2002,jang_theory_2008,trushechkin_calculation_2019}).

The weak coupling interactions are described by molecular quantum electrodynamics (QED), which can be most easily understood using Feynman diagrammatic techniques. This is a unified, fully quantum theory, which produces the $R^{-6}$ and $R^{-2}$ dependence of the short- and long-range interactions respectively as limiting cases \cite{andrews_unified_1989,andrews_virtual_2004,jones_resonance_2019}.

\begin{figure}
\includegraphics[width=\columnwidth]{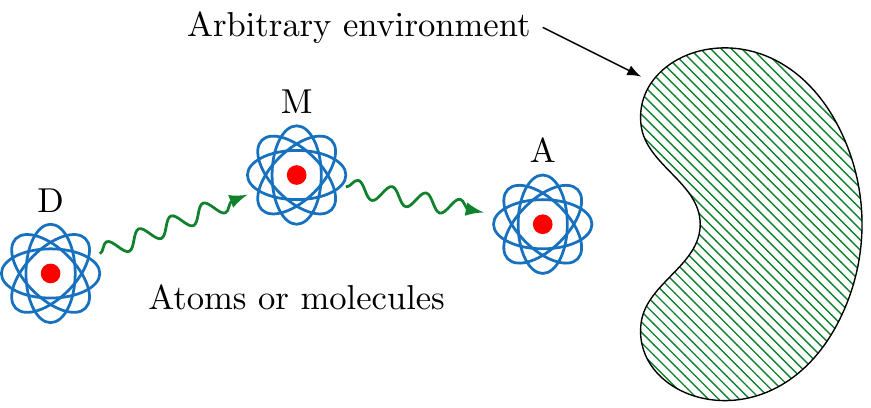}
  \caption{System of three atoms or molecules (donor, mediator and acceptor) in the presence of an arbitrary external environment.}
  \label{fig:schematic1}
\end{figure}

Describing the interaction between a single two-level donor dipole and similar acceptor is relatively straightforward, even in the presence of a background environment \cite{dung_intermolecular_2002,weeraddanaControllingResonanceEnergy2017}. However, the complexity increases significantly with the addition of other levels, additional atoms/molecules \cite{salam_near-zone_2019,craig_third-body_1989,daniels_electronic_2002,salam_mediation_2012,andrews_quantum_2001,bennett_virtual_2019} and higher multipole moments \cite{salam_general_2005}. Calculations which account for a third interacting body have only been done for the simplest case of a vacuum environment --- here we will generalise these to arbitrary environments. In order to describe energy transfer within an external environment (see figure \ref{fig:schematic1}), the molecular QED framework can be combined with macroscopic QED \cite{buhmann_dispersion_2013,gruner_correlation_1995}, allowing interactions in the presence of arbitrarily-shaped dispersive and absorptive media to be described. 

In this paper, a three-body system of a donor, acceptor and a polarisable mediator is studied within macroscopic QED. In section \ref{sec:Hamiltonians}, we will use canonical perturbation theory to eliminate some of the computational complexity arising from the presence of the third body, and in section \ref{sec:Derivation}, use macroscopic QED to model the effects of the external environment. Similar calculations have been carried out for three-body ICD in a vacuum by considering the mediator as part of the environment of the two-body system \cite{bennett_virtual_2019}, but this makes it awkward to extend the calculations to complex geometries. We obtain a general formula for the rate of three-body resonance energy transfer in an arbitrary environment, and in section \ref{sec:Results} we apply this to a situation of experimental interest, namely a dimer trapped near a surface controlled by a distant mediating agent (e.g. \cite{tomasi_coherent_2019}).

\section{Hamiltonians} \label{sec:Hamiltonians}
\begin{figure}
\includegraphics[width=0.62\columnwidth]{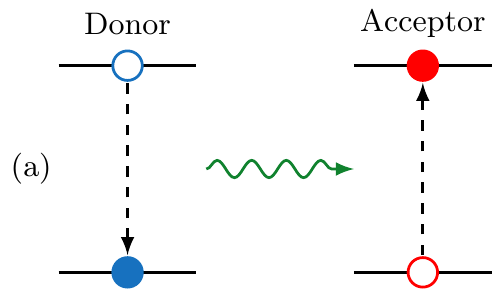}
\includegraphics[width=\columnwidth]{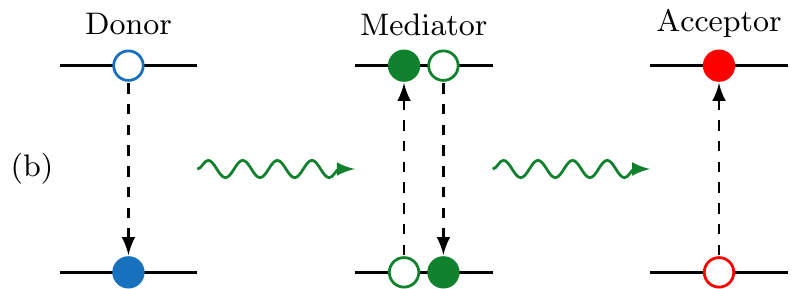}
  \caption{System of three two-level atoms/molecules transmitting energy through the electromagnetic field due to resonance energy transfer. The donor begins in an excited energy state, the acceptor in the ground state and the mediator in its lower state. (a) Direct interaction. Energy is emitted from the donor, is transmitted through the field and absorbed by the acceptor which excites. (b) Mediated interaction. Energy emitted from the donor is absorbed by the mediator causing it to become temporarily excited. The mediator releases this energy again and it is absorbed by the acceptor which then becomes excited.}
\label{fig:levels}
\end{figure}

We consider a system of a donor, acceptor and mediator, as seen in Fig.~\ref{fig:schematic1}. Energy from the donor is released and transferred to the acceptor, either directly (Fig.~\ref{fig:levels}a) or via the mediator (Fig.~\ref{fig:levels}b).
We model this via the following Hamiltonian:
\begin{equation}
H=H_0+ H^{\mathrm{A}}_{\mathrm{int}}+ H^{\mathrm{D}}_{\mathrm{int}}+ H^{\mathrm{M}}_{\mathrm{int}},
\end{equation}
where
\begin{equation}
 H_0=H_{\mathrm{rad}}+H^{\mathrm{A}}_{\mathrm{mol}}+H^{\mathrm{D}}_{\mathrm{mol}}+H^{\mathrm{M}}_{\mathrm{mol}},
\end{equation}
$H_{\mathrm{rad}}$ is the Hamiltonian of the radiation field, $H^\xi_\mathrm{mol}$ is the Hamiltonian of the molecule $\xi$ for which we assume that the eigenstates are known, and
\begin{equation}
 H^\xi_{\mathrm{int}}=-\bm{\hat{d}}_\xi \cdot \bm{\hat{E}}(\bm{r}_\xi),
 \label{eq:InteractionHamiltonian}
\end{equation}
where  $\hat{\bm{d}}_\xi$ is the transition dipole moment of molecule $\xi$, and $\hat{\bm{E}}(\bm{r}_\xi)$ is the electric field at the position, $\bm{r}$, of the molecule $\xi$.

The initial and final states of the system are:
\begin{equation}
\ket{i}=\ket{e_{\mathrm{D}},s_{\mathrm{M}},g_{\mathrm{A}};0}, \qquad
\ket{f}=\ket{g_{\mathrm{D}},s_{\mathrm{M}},e_{\mathrm{A}};0},
\end{equation}
where $g_{\mathrm{D}}(g_{\mathrm{A}})$ denotes the ground state of the donor (acceptor), $e_{\mathrm{D}}(e_{\mathrm{A}})$ the excited state of the donor (acceptor), $s_{\mathrm{M}}$ is an arbitrary state of the mediator and $0$ is the ground state of the electromagnetic field. We need to take into account the mediated interaction between the donor and acceptor, which involves all three molecules. Figure \ref{fig:feynmann}a shows the resonant interaction, where the excitations are transmitted through the field as a result of molecular relaxation from excited to ground states. Likewise, we must also consider the other time-orderings that lead to off-resonant contributions (where molecular excitations are accompanied by emission of a photon), and the half-resonant contributions (where one set of interacting molecules are excited while the photon is transmitted and the other set are in their ground states). All of these contributions have to be considered when calculating rates for the whole process, which confusingly is also known as resonant energy transfer (RET). Figure \ref{fig:feynmann}a shows us that this process would involve four emission/absorption events, meaning that fourth order perturbation theory would be required. The complexity of such a calculation means that it is useful to simplify the Hamiltonian as much as possible before proceeding.

\subsection{Reducing the order of perturbation theory} \label{subsec:Reducing order}
\begin{figure}
\includegraphics[width=0.49\columnwidth]{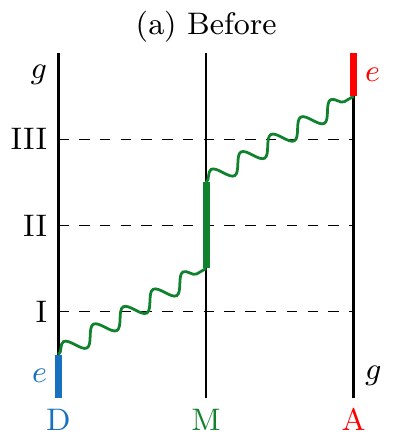}
\includegraphics[width=0.48\columnwidth]{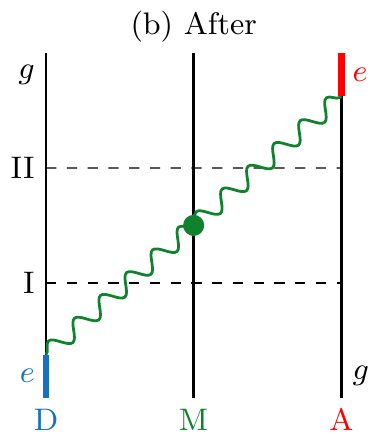}
\caption{The 3-body resonant interaction diagrams (a) before and (b) after the reduction of the order of perturbation theory.}
\label{fig:feynmann}
\end{figure}

The fourth-order calculation consists of contributions from four one-photon vertices. This can be simplified by `collapsing' the two one-photon vertices at the mediator into one two-photon vertex, therefore lowering the perturbation theory required to third order (see Fig.~\ref{fig:feynmann}) as applied at lower orders in \cite{passante_radiation-molecule_1998}.

We are aiming to create a new effective Hamiltonian that encodes the information for the two one-photon interactions at the mediator into one term, so this new coupling term will be second order in the electric dipole moment. To this end, we consider just the interaction at the mediator and not at the donor and acceptor, and perform a unitary transformation on the Hamiltonian,
\begin{align}
H^{\mathrm{M}}_{\mathrm{new}}=&e^{i  S} H^{\mathrm{M}} e^{-i  S} =\sum_{n=0}^{\infty} \frac{1}{n!} \comm{i S}{\comm{i S}{...H}} \notag
\\ =&H_0+ H^{\mathrm{M}}_{\mathrm{int}}+ \comm{iS}{H_0} + \comm{i  S}{H^{\mathrm{M}}_{\mathrm{int}}} \notag
\\&+  \frac{1}{2} \comm{i S}{\comm{i S}{H_0}} +...\quad,
\end{align}
where $S$ is a generator that is assumed later to be first order in the electric dipole moment and we have made use of the Baker–Campbell–Hausdorff formula. We seek a Hamiltonian of second order in $H^{\mathrm{M}}_{\mathrm{int}}$ (and thereby $\hat{\bm{d}}_M$), so we eliminate the first order $H^{\mathrm{M}}_{\mathrm{int}}$ term by choosing $\comm{i  S}{H_0} = - H^{\mathrm{M}}_{\mathrm{int}}$. This leaves, up to second order in the electric dipole moment,
\begin{equation}
 H^{\mathrm{M}}_{\mathrm{new}}=H_0+ \frac{1}{2} \comm{i  S}{H^{\mathrm{M}}_{\mathrm{int}}}
\end{equation}
where, from our chosen definition of the generator, for initial state $\ket{M}$ and final state $\ket{N}$ we have
\begin{equation}
\bra{N} iS \ket{M} = \frac{\bra{N} H^{\mathrm{M}}_{\mathrm{int}} \ket{M}}{E_N-E_M}.
\end{equation}

We now calculate the expectation value of the new seconder order interaction term using the definition of generator $S$ given above;
\begin{align}
\frac{1}{2}& \bra{N} \comm{i S}{H^{\mathrm{M}}_{\mathrm{int}}} \ket{M} \notag
\\ = -& \frac{1}{2} \sum_{\mathrm{I}} \bra{N} H^{\mathrm{M}}_{\mathrm{int}} \ket{I}\bra{I}  H^{\mathrm{M}}_{\mathrm{int}}\ket{M}
\left[\frac {1}{E_{\mathrm{I}}-E_{\mathrm{N}}} + \frac{1}{E_{\mathrm{I}}-E_{\mathrm{M}}} \right] \notag
\\  = -& \frac{1}{2} \sum_r \bra{N} H^{\mathrm{M}}_{\mathrm{int}} H^{\mathrm{M}}_{\mathrm{int}}\ket{M} \notag
\\ \times& \left[\frac {1}{E_{\mathrm{rs}}+\hbar cp} + \frac{1}{E_{\mathrm{rs}}-\hbar cp} + \frac{1}{E_{\mathrm{rs}}+\hbar cp'} + \frac{1}{E_{\mathrm{rs}}-\hbar cp'}\right],
\end{align}
where $E_{rs}$ is the transition energy of the mediator going from the excited $r$ state to its lower $s$ state.

\begin{figure}
\includegraphics[width=0.8\columnwidth]{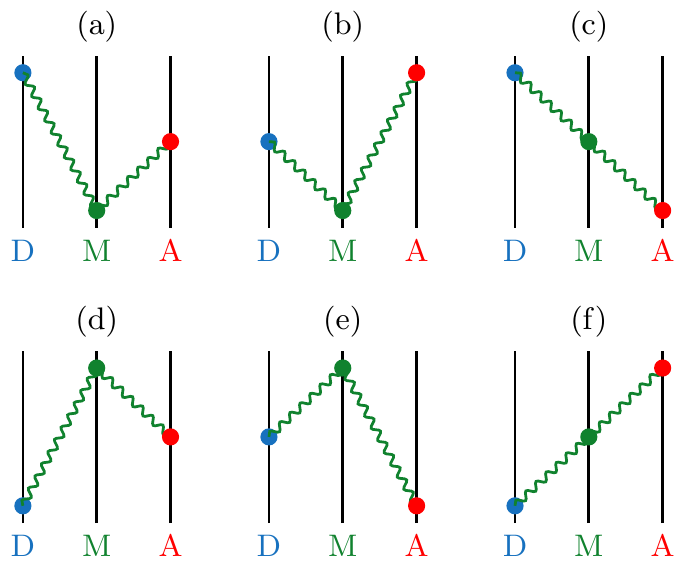}
  \caption{The six time ordered diagrams for three-body RET once the two one-photon vertices have been collapsed into one two-photon vertex.}
  \label{fig:Simplertimeorderings1}
\end{figure}

Since in three-body RET, the mediator responds at the frequency of the donor decay transition \cite{salam_near-zone_2019}, we can replace $\hbar cp \mapsto \hbar ck = E_{\mathrm{eg}}$ and $\hbar cp' \mapsto -\hbar ck = -E_{\mathrm{eg}}$, giving
\begin{multline}
\frac{1}{2} \bra{N}  \comm{i S}{H^{\mathrm{M}}_{\mathrm{int}}} \ket{M}
 \\ = - \sum_r \bra{N} H^{\mathrm{M}}_{\mathrm{int}} H^{\mathrm{M}}_{\mathrm{int}}\ket{M} \left[\frac {1}{E_{\mathrm{rs}}+E_{\mathrm{eg}}} + \frac{1}{E_{\mathrm{rs}}-E_{\mathrm{eg}}}\right].
\end{multline}
We can therefore define a new coupling term as given below:
\begin{equation}
H_2=\frac{1}{2}\comm{i S}{H^{\mathrm{M}}_{\mathrm{int}}} = -\sum_{p,p'}  \alpha^{\mathrm{M}}_{ij}(k) \hat{E}_i(\bm{r}_{\mathrm{M}},p) \\ \hat{E}_j(\bm{r}_{\mathrm{M}},p')
\label{eq:H2}
\end{equation}
where
\begin{equation}
\alpha^{\mathrm{M}}_{ij}(k) = \sum_{r} |\bm{\hat{d}}^{\mathrm{M}}_{\mathrm{sr}} |^2
\left[  \frac {1 } {E_{\mathrm{rs}} + \hbar ck}  + \frac{1} {E_{\mathrm{rs}} - \hbar ck}\right]
\end{equation}
is identified as the dynamic polarizability of the mediator. This means that when considering 3-body RET, instead of fourth order perturbation theory being required, now only third order is needed. This reduces the usual 24 time-ordered diagrams required for this calculation to just six, as shown in figure \ref{fig:Simplertimeorderings1}. This process is, of course, equivalent to using a polarisability-based Hamiltonian, as done in \cite{salam_bridge-mediated_2021} for example.

We now go one step further and use the same techniques to create an effective fourth order term that describes the entire interaction including all three bodies. We begin with the new Hamiltonian,
\begin{equation}
    H_{\mathrm{new}}=H_0+ H_1+ H_2,
\end{equation}
where $H_2$ is the defined as in equation \eqref{eq:H2} and, for convenience, the acceptor and donor interaction terms have been condensed into a single term, $H_1=H^{\mathrm{A}}_{\mathrm{int}}+ H^{\mathrm{D}}_{\mathrm{int}}$, which is linear in the electric dipole moment. Introducing the dimensionless constant $\lambda$ which is proportional to the electric dipole moment, so that 
\begin{equation}
H_{\mathrm{new}}=H_0+ \lambda H_1 + \lambda^2 H_2,
\end{equation}
we perform a series of unitary transformations:
\begin{equation}
  H_{\mathrm{new}}^{(1)}=e^{i \lambda^3 S_3}e^{i \lambda^2 S_2}e^{i \lambda S_1} H_{\mathrm{new}} e^{-i \lambda S_1}e^{-i \lambda^2 S_2}e^{-i \lambda^3 S_3}
\end{equation}
with generators defined as follows:
\begin{align}
  \nonumber \comm{i  S_1}{H_0}&=- H_1 \notag \\
  \nonumber \comm{iS_2}{H_0}&=-H_2 \notag \\
  \comm{iS_3}{H_0}&=-\comm{iS_1}{H_2}-\comm{iS_1}{\comm{iS_1}{H_1}}.
\end{align}
These are chosen such that any donor-mediator-acceptor interactions are eliminated to order $\lambda^3$.

After dropping any second order terms that do not contain donor and acceptor contributions, and any fourth order terms that do not contain mediator contributions, we arrive at our new interaction Hamiltonian, the mediator-dependent parts of which are fourth order in the electric dipole moment as required;
\begin{equation}\label{SHamiltonian}
H_{\mathrm{int}} = \frac{1}{2}\left(\comm{iS_1}{H_1} + \comm{iS_1}{\comm{iS_1}{H_2}} +  \comm{iS_2}{\comm{iS_1}{H_1}} \right).
\end{equation}
This will form the basis of our perturbative treatment of mediated resonance energy transfer.

\subsection{Perturbation theory} \label{subsec:PT}
We now model our three-body system using the Hamiltonian \eqref{SHamiltonian}, and perform perturbation theory to find the matrix element of the interaction that will eventually lead to a RET rate via Fermi's golden rule. Considering both the direct and indirect interactions, we find the matrix element to be
\begin{align}
&M_{fi}=\sum_{p,p',k} \bra{f} \Biggl[
\frac{H_{\mathrm{A}} H_{\mathrm{D}} }{\hbar cp-E_{\mathrm{eg}}} -\frac{H_{\mathrm{D}} H_{\mathrm{A}}}{\hbar cp+E_{\mathrm{eg}}} \notag \\ &+\frac{H_{\mathrm{D}} H_{\mathrm{A}} H_2}{(\hbar cp+E_{\mathrm{eg}})(\hbar cp'-E_{\mathrm{eg}})} 
+ \frac{H_2 H_{\mathrm{D}} H_{\mathrm{A}}}{(\hbar \notag cp-E_{\mathrm{eg}})(\hbar cp'+E_{\mathrm{eg}})}  
\\ &+\frac{H_{\mathrm{D}} H_2 H_{\mathrm{A}}}{(\hbar cp+E_{\mathrm{eg}})(\hbar cp'+E_{\mathrm{eg}})} + \frac{H_{\mathrm{A}} H_2 H_{\mathrm{D}}}{(\hbar cp-E_{\mathrm{eg}})(\hbar cp'-E_{\mathrm{eg}})}
\Biggr] \ket{i}
\label{eq:Mfi}
\end{align}
where, using Eq.~\eqref{eq:InteractionHamiltonian}, 
\begin{align}
   H_{\mathrm{D}}&\equiv H_{\mathrm{D}}^{\mathrm{int}} =-\bm{\hat{d}}_{\mathrm{A}} \cdot
   \bm{\hat{E}}(\bm{r}_{\mathrm{D}}) \notag \\
   H_{\mathrm{A}}&\equiv H_{\mathrm{D}}^{\mathrm{int}}=-\bm{\hat{d}}_{\mathrm{A}} \cdot \bm{\hat{E}}(\bm{r}_{\mathrm{A}}) \notag \\
   H_2&=- \alpha^{\mathrm{M}}_{ij}(k) \hat{E}_i(\bm{r}_{\mathrm{M}},p) \hat{E}_j(\bm{r}_{\mathrm{M}},p').
\end{align}
This matrix element contains the information from all of the different time orderings, meaning we no longer have any explicit intermediate states. As a result, the order of perturbation theory is further reduced to first order. The first two terms in \eqref{eq:Mfi} are the direct (two-body) interaction terms, the first corresponding to the resonant time ordering and the second the off-resonant. The other terms describe different time orderings of the mediated interaction. The third and fourth terms are the half-resonant contributions, where the third term corresponds to (d) and (e) in Fig.~\ref{fig:Simplertimeorderings1} and the fourth term to (a) and (b). The fifth term describes the completely off-resonant interaction, shown in (c), and the sixth is the completely resonant interaction, (f). We can then use this matrix element to calculate the rate of interaction.

\section{Derivation of the Rate} \label{sec:Derivation}
So far, we have not considered the environment that the 3-body system is in. To do this, we employ macroscopic QED \cite{dung_three-dimensional_1998,gruner_correlation_1995}, which introduces macroscopic objects into the quantum description. This means that the effects of an environment near the system can be accounted for more readily than the atomistic approach in \cite{juzeliifmmode_baruelse_ufinas_quantum_1994}, since the macroscopic media can be described by their effective properties, such as overall permittivity and permeability. This environment can include arbitrarily shaped, dispersing, and absorbing material bodies. 

In macroscopic QED, the electric field is expressed as;
\begin{equation}
\bm{\hat{E}}(\bm{r}) =  \sum_{\lambda} \int_0^\infty d\omega \int d^3\bm{r}'
 \bm{G}_{\lambda}(\bm{r},\bm{r}',\omega) \cdot \bm{\hat{f}}_{\lambda}(\bm{r}',\omega)
 +H.c,
\end{equation}
where $\hat{\bm{f}}_{\lambda}(\bm{r},\omega)$ is an annihilation operator for a polaritonic excitation at position $\bm{r}$ and with frequency $\omega$, and its Hermitian conjugate is the corresponding creation operator. These operators obey bosonic commutation relations
\begin{align}
    \left[ \bm{\hat{f}}_{\lambda}(\bm{r},\omega),\bm{\hat{f}}_{\lambda}(\bm{r}',\omega')\right]&=\left[ \bm{\hat{f}}^\dagger_{\lambda}(\bm{r},\omega),\bm{\hat{f}}^\dagger_{\lambda}(\bm{r}',\omega')\right]=0 
    \\ \left[ \bm{\hat{f}}_{\lambda}(\bm{r},\omega),\bm{\hat{f}}^\dagger_{\lambda}(\bm{r}',\omega')\right] &= \bm{\delta}(\bm{r}-\bm{r}') \delta(\omega-\omega').
\end{align}
where $\bm{\delta}(\bm{r}-\bm{r}')=\mathrm{diag}(1,1,1)\delta(\bm{r}-\bm{r}')$.
The matrix $\bm{G}_\lambda$ obeys the completeness relation
\begin{multline}
\sum_{\lambda = e,m} \int d^3 \bm{s}  \bm{G}_{\lambda}(\bm{r},\bm{s},\omega) \cdot \bm{G}_{\lambda}^{*T}(\bm{r}',\bm{s},\omega)
\\ = \frac{\hbar \mu_0}{\pi} \omega^2 \Im \bm{G}(\bm{r},\bm{r}',\omega),
\label{eq:completeness}
\end{multline}
where $\bm{G}$ satisfies
\begin{equation}
\left[\nabla \times \frac{1}{\mu (\bm{r},\omega)} \nabla \times -\frac{\omega^2}{c^2} \times \varepsilon (\bm{r},\omega) \right] \bm{G}(\bm{r},\bm{r'},\omega)=\delta(\bm{r}-\bm{r'}).
\end{equation}
Known as the Green's tensor or Green's dyadic, $\bm{G}$ encodes all of the information about the environment, including different geometries \cite{tai1994dyadic}. In particular, $\bm{G}(\bm{r},\bm{r}',\omega)$ describes an excitation that is propagating out from point $\bm{r}'$ and then being observed at point $\bm{r}$. 

By making use of the completeness relation \eqref{eq:completeness}, we can write each of the terms in the matrix elements as frequency integrals of the form;
\begin{equation}
   \int^{\infty}_0 d\omega \frac{\omega ^2 \Im \bm{G}(\bm{r},\bm{r}',\omega)}{\omega_{\mathrm{D}} \pm \omega}.
\end{equation}
For example, the first term of the matrix element becomes:
\begin{align}
\sum_{k} &\bra{f} \frac{H_{\mathrm{A}} H_{\mathrm{D}} }{E_{\mathrm{eg}}-\hbar ck} \ket{i} \notag
\\ &= \frac{\mu_0}{\pi}  d_{A_i}^*
\int^{\infty}_0 d\omega \frac{\omega^2 \Im G_{ij}(\bm{r}_{\mathrm{A}},\bm{r}_{\mathrm{D}},\omega)}{\omega-\omega_{\mathrm{D}}}  d_{D_j},
\end{align}
where we have defined
\begin{align}
\hbar cp&=\hbar \omega,  & E_{\mathrm{eg}} &= \hbar \omega_{\mathrm{D}}
\\ \bm{d} &\equiv \bra{g}\bm{\hat{d}}\ket{e},  & \bm{d^*} &\equiv \bra{e}\bm{\hat{d}}\ket{g}.
\end{align}

Since the frequency integrals have real axis poles, we let the eigenenergies of the atom take on a small imaginary part, $\epsilon$. This means that the poles become $\pm (\omega_{\mathrm{D}}+i\epsilon)$, and we then evaluate the integrals by closing the contour in the upper half-plane \cite{jenkins_quantum_2004,jones_resonance_2019}. We find that;
\begin{multline}
   \lim_{\epsilon \rightarrow 0+} \int^{\infty}_0 d\omega \frac{\omega ^2 \Im \bm{G}(\bm{r},\bm{r}',\omega)}{\omega_{\mathrm{D}} + \omega +i\epsilon}
   \\ =  - \int^{\infty}_0 d\xi \bm{G}(\bm{r},\bm{r}',i\xi) \xi^2
\frac{\omega_{\mathrm{D}}}{\omega_{\mathrm{D}}^2+\xi^2}
\end{multline}
and
\begin{multline}
   \lim_{\epsilon \rightarrow 0+} \int^{\infty}_0 d\omega \frac{\omega ^2 \Im \bm{G}(\bm{r},\bm{r}',\omega)}{\omega_{\mathrm{D}} - \omega +i\epsilon}
   \\ =  - \int^{\infty}_0 d\xi \bm{G}(\bm{r},\bm{r}',i\xi) \xi^2
\frac{\omega_{\mathrm{D}}}{\omega_{\mathrm{D}}^2+\xi^2}
- \pi \omega_{\mathrm{D}}^2 \bm{G}(\bm{r},\bm{r}',\omega_{\mathrm{D}}).
\end{multline}
Applying this method to each of the terms in the matrix element, and 
assuming that all media involved are reciprocal, so that Lorentz reciprocity $\bm{G}^T(\bm{r},\bm{r}',\omega)=\bm{G}(\bm{r}',\bm{r},\omega)$ can be used, we find that the surviving term for the direct contribution (see Fig.~\ref{fig:levels}a) to the matrix element is
\begin{equation}
M_{fi}^{\mathrm{dir}}=\mu_0 \omega_{\mathrm{D}}^2 d_{A_i}^* {G}_{ij}(\bm{r}_{\mathrm{A}},\bm{r}_{\mathrm{D}},\omega_{\mathrm{D}})  d_{D_j}
\end{equation}
and for the indirect contribution (see Fig.~\ref{fig:levels}b)
\begin{multline}
M_{fi}^{\mathrm{indir}} =- \mu_0^2 \omega_{\mathrm{D}}^4  d^*_{A_i}
  {G}_{ij}(\bm{r}_{\mathrm{A}},\bm{r}_{\mathrm{M}},\omega_{\mathrm{D}})
\\ \times \alpha^{\mathrm{M}}_{jk}(k)
 {G}_{kl}(\bm{r}_{\mathrm{M}},\bm{r}_{\mathrm{D}},\omega_{\mathrm{D}}) d_{D_l}.
\end{multline}

Summing these contributions to find the total matrix element $M_{fi}=M_{fi}^{\mathrm{dir}}+M_{fi}^{\mathrm{indir}}$, we calculate the rate from Fermi's Golden rule as
\begin{align}
\Gamma &= \sum_f \frac{2\pi}{\hbar}|M_{fi}|^2 \delta(E_{i}-E_f) \notag
\\&=\frac{2\pi \mu_0^2 \omega_{\mathrm{D}}^4 }{\hbar}
|\bm{d}_{\mathrm{A}}^* \cdot \big[ \bm{G}(\bm{r}_{\mathrm{A}},\bm{r}_{\mathrm{D}},\omega_{\mathrm{D}}) \notag
\\ +\mu_0 &\omega_{\mathrm{D}}^2 \bm{G}(\bm{r}_{\mathrm{A}},\bm{r}_{\mathrm{M}},\omega_{\mathrm{D}}) \cdot
\bm{\alpha}^{\mathrm{M}}(k) \cdot \bm{G}(\bm{r}_{\mathrm{M}},\bm{r}_{\mathrm{D}},\omega_{\mathrm{D}}) \big] \cdot \bm{d}_{\mathrm{D}}|^2.
\label{eq:rate}
\end{align}
which reduces to the three-body ICD formula found in \cite{bennett_virtual_2019} if the Green's tensor is replaced by its vacuum counterpart and the transition dipole moment of the acceptor is re-expressed in terms of an ionisation cross section. Equation \eqref{eq:rate} is the main result of our work, describing resonant energy transfer mediated by a third polarisable body in the presence of an arbitrary environment. The first term of \eqref{eq:rate} describes the direct interaction between the donor and acceptor, where the field propagates from the donor at position $\bm{r}_\mathrm{D}$ and is observed at the acceptor at $\bm{r}_\mathrm{A}$, therefore corresponding to the resonant interaction. The second term describes the mediated interaction, where the field propagates from the donor to the mediator, and then from the mediator to the acceptor. We can see therefore that while all diagrams in Fig.~\ref{fig:Simplertimeorderings1} contribute, the result is what one would have obtained from the pole contributions to the resonant diagram only (as was assumed without rigorous justification in \cite{bennett_virtual_2019}), with all the remaining diagrams serving to cancel the non-pole parts of this.

\section{Example application; external control of a molecular dimer} \label{sec:Results}
Formula \eqref{eq:rate} allows calculation of the rate of mediated energy transfer in an arbitrary external environment. As a proof-of-concept, we demonstrate the use of the formula for the simplest inhomogeneous environment, namely a semi-infinite half-space. We will specialise to some asymptotic distance regimes in order to be able to write down simple Green's tensors in position space (i.e. without using an angular spectrum representation), verifying these against full results at the end. However we emphasise that the formula \eqref{eq:rate} is applicable to any external environment, and could be used to calculate interactions within far more complex systems using numerically calculated Green's tensors. These could include proteins and other biological systems \cite{PISTON2007407}, in which the dipole moments are often randomly oriented, necessitating calculating the rate averaged over all possible dipole alignments. The procedure for this is well-known (see, for example \cite{andrews_three-dimensional_1989}), in particular it amounts to making the replacement for the outer product of a dipole moment with itself.
\begin{equation}
\bm{d}_{\mathrm{A/D}}^* \otimes \bm{d}_{\mathrm{A/D}} \to \frac{1}{3} |\bm{d}_{\mathrm{A/D}}|^2 \mathbb{I},
\end{equation}
where $\otimes$ denotes the outer product and $\mathbb{I}$ is the $3\times 3$ identity matrix. Multiplying out Eq.~\eqref{eq:rate}, applying this rule and again taking advantage of Lorentz reciprocity, we find;
\begin{align}
\Gamma^\mathrm{iso} &=\frac{2\pi \mu_0^2 \omega_{\mathrm{D}}^4 }{9\hbar}
|\bm{d}_{\mathrm{A}}|^2|\bm{d}_{\mathrm{D}}|^2\notag \\
&\qquad \times \text{Tr}[\bm{F}(\bm{r}_{\mathrm{A}},\bm{r}_{\mathrm{M}},\bm{r}_{\mathrm{D}}) \cdot \bm{F}^*(\bm{r}_{\mathrm{D}},\bm{r}_{\mathrm{M}},\bm{r}_{\mathrm{A}})],
\label{eq:rateIso}
\end{align}
where
\begin{align}
    \bm{F}(\bm{r}_{\mathrm{A}},&\bm{r}_{\mathrm{M}},\bm{r}_{\mathrm{D}})\equiv\bm{G}(\bm{r}_{\mathrm{A}},\bm{r}_{\mathrm{D}},\omega_{\mathrm{D}})\notag \\
    &+\mu_0 \alpha^\mathrm{M} \omega_{\mathrm{D}}^2 \bm{G}(\bm{r}_{\mathrm{A}},\bm{r}_{\mathrm{M}},\omega_{\mathrm{D}}) \cdot \bm{G}(\bm{r}_{\mathrm{M}},\bm{r}_{\mathrm{D}},\omega_{\mathrm{D}}). \label{eq:F}
\end{align}
and we have let $\bm{\alpha}^{\mathrm{M}}\to{\alpha}^{\mathrm{M}}\mathbb{I}$ in order to effect an isotropic mediator.

\begin{figure}
\includegraphics[width=0.9\linewidth]{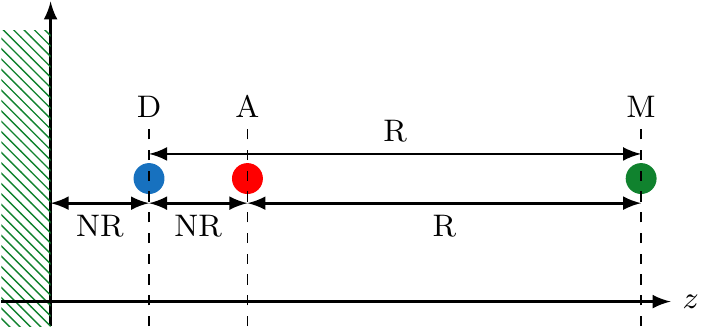}
  \caption{Colinear system made up of three bodies and a semi-infinite dielectric half-space. The donor and acceptor are assumed close enough together and to the surface to apply the non-retarded (NR) limit to their direct interaction, and the mediator is assumed far enough away from both that the retarded (R) limit can be applied.}
  \label{fig:colinearHS}
\end{figure}

For simplicity we will initially look at a colinear system, where all three bodies are positioned along the $z$-axis, as shown in Fig.~\ref{fig:colinearHS}. Inspired by schemes that aim to control enhancement of light-harvesting efficiency (e.g. \cite{tomasi_coherent_2019}), we choose to look at the case where the donor and acceptor are positioned very close to each other and also to the half-space, but with a distant mediator. This means that we can make further simplifications, namely that the donor and  acceptor are close enough to each other and the surface that the environment-dependent direct interaction between them is in the non-retarded limit, where the intermolecular distance between the donor and acceptor is significantly less that the characteristic wavelength of the transition. The other limit we impose is that the mediator is far enough away from the donor, acceptor and surface that the opposite, retarded limit can be used there. 

We can therefore write an approximate rate $\Gamma^\mathrm{iso}_\mathrm{CL}$ for our colinear system as;
\begin{align}
\Gamma^\mathrm{iso}_\mathrm{CL} &=\frac{2\pi \mu_0^2 \omega_{\mathrm{D}}^4 }{9\hbar}
|\bm{d}_{\mathrm{A}}|^2|\bm{d}_{\mathrm{D}}|^2\notag \\
&\qquad \times \text{Tr}[\bm{F}_\mathrm{CL}(\bm{r}_{\mathrm{A}},\bm{r}_{\mathrm{M}},\bm{r}_{\mathrm{D}}) \cdot \bm{F}_\mathrm{CL}^*(\bm{r}_{\mathrm{D}},\bm{r}_{\mathrm{M}},\bm{r}_{\mathrm{A}})],
\label{eq:rateIsoCL}
\end{align}
where
\begin{align}
    \bm{F}_\mathrm{CL}&(\bm{r}_{\mathrm{A}},\bm{r}_{\mathrm{M}},\bm{r}_{\mathrm{D}})\equiv\bm{G}_\mathrm{NR}(\bm{r}_{\mathrm{A}},\bm{r}_{\mathrm{D}},\omega_{\mathrm{D}})\notag \\
    &+\mu_0 \alpha^\mathrm{M} \omega_{\mathrm{D}}^2 \bm{G}_\mathrm{R}(\bm{r}_{\mathrm{A}},\bm{r}_{\mathrm{M}},\omega_{\mathrm{D}}) \cdot \bm{G}_\mathrm{R}(\bm{r}_{\mathrm{M}},\bm{r}_{\mathrm{D}},\omega_{\mathrm{D}}). \label{eq:FLimits}
\end{align}

In order to explicitly calculate the approximate colinear rate $\Gamma^\mathrm{iso}_\mathrm{CL}$ defined in \eqref{eq:rateIsoCL}, we note that Green's tensors in inhomogenous environments can in general be split into the sum of a translation-invariant bulk part $\bm{G}^{(0)}$ and a `scattering' part $\bm{G}^{(1)}$, so that each of the Green's tensors shown in \eqref{eq:FLimits} are;
\begin{equation}
    \bm{G}_{\mathrm{NR/R}}(\bm{r},\bm{r}',\omega)= \bm{G}_{\mathrm{NR/R}}^{(0)}(\bm{r},\bm{r}',\omega)+ \bm{G}_{\mathrm{NR/R}}^{(1)}(\bm{r},\bm{r}',\omega)
\end{equation}
For our chosen system (dielectric material in the region $z<0$, vacuum otherwise), the bulk part of the Green's tensor is that of the vacuum which is known analytically in closed form (see, e.g. \cite{buhmann_dispersion_2013} or \cite{tai1994dyadic}), but due to the physical system we have chosen we will only quote its short and long distance forms. In the non-retarded limit (with non-coincident position arguments $\bm{r}\neq \bm{r}'$), the vacuum Green's tensor reads \cite{buhmann_dispersion_2013}
\begin{equation}
    \bm{G}_\mathrm{NR}^{(0)}(\bm{r},\bm{r}',\omega)=- \frac{c^2 e^{i\omega\rho/c}}{4\pi \omega^2 \rho^3} (\mathbb{I}-3 \bm{e}_{\rho} \otimes\bm{e}_{\rho} ),
    \label{eq:G0NR}
\end{equation}
where $\bm{\rho}=\bm{r}-\bm{r}'$, $\rho=|\bm{\rho}|$ and $\bm{e}_{\rho}=\bm{\rho}/\rho$. In the retarded limit it becomes
\begin{equation}
    \bm{G}_\mathrm{R}^{(0)}(\bm{r},\bm{r}',\omega)=- \frac{c^2 e^{i\omega\rho/c}}{4\pi \rho} ( \mathbb{I}- \bm{e}_{\rho} \otimes\bm{e}_{\rho} ).
    \label{eq:G0R}
\end{equation}

Moving onto the scattering contribution, we note that the full expression of the scattering Green's tensor for a general dielectric half-space can only be written in terms of Fourier-transformed quantities (see, e.g. \cite{buhmann_dispersion_2013} or \cite{tai1994dyadic}, necessitating one or more frequency integrals before results can be obtained. Fortunately, due to our choice of physical situation we can again use the short and long distance special cases, which can be written as simple expressions in position space. The non-retarded limit of the scattering Green's tensor on the positive $z$ axis for an environment containing a non-magnetic (relative permeability of unity), semi-infinite half-space in the region $z<0$ is \cite{buhmann_dispersion_2012};
\begin{equation}
\bm{G}^{(1)}_\mathrm{NR}(\bm{r},\bm{r}',\omega)= \frac{c^2}{4\pi\omega^2(z+z')^3} r_\mathrm{NR} \begin{pmatrix} 1&0&0\\0&1&0\\0&0&2 \end{pmatrix}
\label{eq:G1NR}
\end{equation}
and the retarded limit is;
\begin{equation}
\bm{G}^{(1)}_\mathrm{R}\bm{r},\bm{r}',\omega)=\frac{e^{i(z+z')\omega/c}}{4\pi (z+z')} r_\mathrm{R} \begin{pmatrix} 1&0&0\\0&1&0\\0&0&0 \end{pmatrix},
\label{eq:G1R}
\end{equation}
where
\begin{align}
    r_\mathrm{NR} &= \frac{\varepsilon(\omega)-1}{\varepsilon(\omega)+1}, & r_\mathrm{R} = \frac{1-\sqrt{\varepsilon(\omega)}}{1-\sqrt{\varepsilon(\omega}},
\end{align}
are the reflection coefficients of the half-space for the non-retarded and retarded limits, $z$ and $z'$ are the $z$ components of the distances $\bm{r}$ and $\bm{r'}$, and $\varepsilon(\omega)$ is the frequency-dependent relative permittivity of the half-space. Equations \eqref{eq:rateIsoCL}-\eqref{eq:G1R} together constitute an analytic formula for the donor-acceptor transfer rate in the situation shown in Fig.~\eqref{fig:colinearHS}. 
Substituting Eqs~\eqref{eq:FLimits}-\eqref{eq:G1R} into \eqref{eq:rateIsoCL} and simplifying, we find
\begin{align}\label{RateFinalISO}
&\Gamma^\mathrm{iso}_\mathrm{CL} = \frac{\mu _0^2c^4 |\bm{d}_{\mathrm{A}}|^2|\bm{d}_{\mathrm{D}}|^2}{18\pi \hbar }\notag\\
&\times \Bigg\{ \left[\frac{r_{\text{NR}}}{\left(z_{\text{A}}+z_{\text{D}}\right)^3}+\frac{1}{\left(z_{\text{A}}-z_{\text{D}}\right)^3}\right]^2
+\frac{|C|^2}{32 \pi ^2 c^4}\Bigg\},
\end{align}
where
\begin{align}
C=\,&4 \pi  c^2 \left(\frac{1}{\left(z_{\text{A}}-z_{\text{D}}\right){}^3}-\frac{r_{\text{NR}}}{\left(z_{\text{A}}+z_{\text{D}}\right){}^3}\right)\notag \\
&-\Bigg\{\frac{\mu _0 \omega _{\text{D}}^4 \alpha _{\text{M}} e^{-{i \omega _{\text{D}} \left(z_{\text{A}}+z_{\text{D}}-2 z_{\text{M}}\right)}/{c}}}{\left(z_{\text{A}}-z_{\text{M}}\right) \left(z_{\text{A}}+z_{\text{M}}\right) \left(z_{\text{M}}-z_{\text{D}}\right) \left(z_{\text{D}}+z_{\text{M}}\right)}\notag \\
&\times \left(r_{\text{R}} \left(z_{\text{A}}-z_{\text{M}}\right) e^{{2 i z_{\text{A}} \omega _{\text{D}}}/{c}}-z_{\text{A}}-z_{\text{M}}\right)\notag\\ &\times \left(r_{\text{R}} \left(z_{\text{M}}-z_{\text{D}}\right) e^{{2 i \omega _{\text{D}} z_{\text{D}}}/{c}}+z_{\text{D}}+z_{\text{M}}\right)\Bigg\}.
\end{align}
with $z_\text{D}<z_\text{A}<z_\text{M}$, as indicated in Fig.~\ref{fig:colinearHS}. In the limit of vanishing mediator polarisibility $\alpha_\text{M}$, all terms depending on the retarded reflection coefficient $r_\mathrm{R}$ vanish (as is expected from the assumption that the mediator is at a retarded distance from all other bodies), and one is left with

\begin{align}
   \Gamma_0 &\equiv \Gamma^\mathrm{iso}_\mathrm{CL}(\alpha_\text{M} \to 0) = \frac{c^4 \mu _0^2 |\bm{d}_{\mathrm{A}}|^2|\bm{d}_{\mathrm{D}}|^2}{36 \pi  \hbar }\Bigg[\frac{3 r_{ \text{NR}}^2}{\left(z_{\text{A}}+z_{\text{D}}\right){}^6}\notag \\
   &+\frac{2 r_{\text{NR}}}{\left(z_{\text{A}}-z_{\text{D}}\right){}^3 \left(z_{\text{A}}+z_{\text{D}}\right){}^3}+\frac{3}{\left(z_{\text{A}}-z_{\text{D}}\right){}^6}\Bigg].\label{GammaIso0}
\end{align}

To the best of our knowledge, this formula for the two-body isotropically-averaged rate near an dielectric interface does not appear anywhere in the literature, the closest known result being the that for oriented (non-random) dipoles near a perfect reflector reported in  Eq.~(20) of \cite{weeraddanaControllingResonanceEnergy2017}. In Appendix \ref{ConsistencyAppendix} we demonstrate that the result in \cite{weeraddanaControllingResonanceEnergy2017} is exactly reproduced by the relevant special case of Eq.~\eqref{eq:rate}. Finally, taking the limit of \eqref{GammaIso0} where the surface becomes transparent ($r_{\text{NR}}\to 0$), one finds;
\begin{align}
   \Gamma(\alpha_\text{M} \to 0,r_{\text{NR}}\to 0) =& \frac{c^4 \mu _0^2 |\bm{d}_{\mathrm{A}}|^2|\bm{d}_{\mathrm{D}}|^2}{12 \pi  \hbar \left(z_{\text{A}}-z_{\text{D}}\right)^6},
\end{align}
in agreement with the well-known result for two-body resonance energy transfer in vacuum \cite{forsterZwischenmolekulareEnergiewanderungUnd1948}

\begin{figure}
\centering
\includegraphics[width=\linewidth]{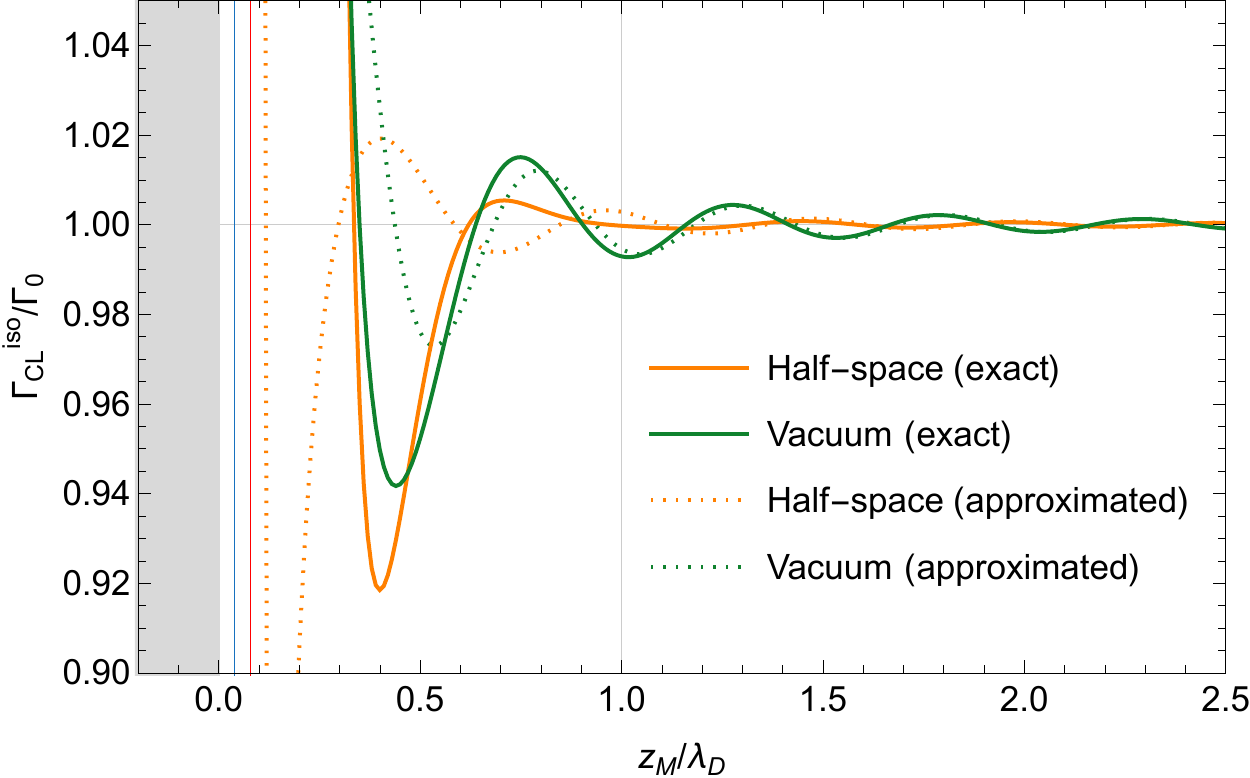}
  \caption{Plot of rate of energy transfer \eqref{RateFinalISO} against mediator position for a system in a vacuum and a system near a half-space (gray) modelled as a perfect reflector (corresponding to $\varepsilon \to \infty$ so that $r_\mathrm{NR}=1=-r_\mathrm{R}$). All rates are normalised to their isotropic two-body rate $\Gamma_0 = \Gamma^\mathrm{iso}_\mathrm{CL}(\alpha^\mathrm{M} \to 0)$ [see Eq.~\eqref{GammaIso0}] in their respective environments, and the mediator position is in units of the donor transition wavelength $\lambda_\mathrm{D}$. The polarizability volume $ \alpha^\mathrm{M}/4\pi \epsilon_0$ of the mediator is chosen as $0.1 \lambda_\text{D}^3$. The surface is positioned at $z=0$, the donor is at $z/\lambda_\mathrm{D}= 0.04$ and the acceptor is at $z/\lambda_\mathrm{D}= 0.08$ as indicated by the blue and red vertical lines and dictated by the imposition of the non-retarded limit in that section of the system. In order for the retarded approximation to hold in its section of the system, the mediator should not be brought nearer than approximately a wavelength away from the donor, acceptor or boundary --- this is indicated by the vertical line.}
  \label{fig:colinearplot1}
\end{figure}

Returning to our three-body, environment modified rate formula \eqref{RateFinalISO}, based on Ref.~\cite{tomasi_coherent_2019} we are particularly interested in how changing the position of the mediator affects the rate of energy transfer between the donor and acceptor. Figure \ref{fig:colinearplot1} shows how the rate of energy transfer changes as the position of the mediator is varied along the $z$-axis in the presence of a half-space, and compares this with a vacuum environment, both using the approximate formula \eqref{eq:rateIsoCL} and for an exact numerical calculation using the full formula \eqref{eq:rateIso} (carried out using the Fourier-transformed Green's tensors found in, for example, appendix B of \cite{buhmann_dispersion_2013}). It is clear from Fig.~\ref{fig:colinearplot1} that the approximations we applied to write down Eq.~\eqref{eq:rateIsoCL} work where they are expected to (mediator significantly more than one wavelength away from donor, acceptor and surface), but fail outside of that. It is interesting to note from Fig.~\ref{fig:colinearplot1} that for this particular situation the effect of the mediator is actually diminished by the presence of the half-space. In other words, when the environment contains a half-space, adding a controllable third body will have a less of effect on the energy transfer rate between the donor and acceptor than if no half-space were present.

\begin{figure}
\centering
\includegraphics[width=\columnwidth]{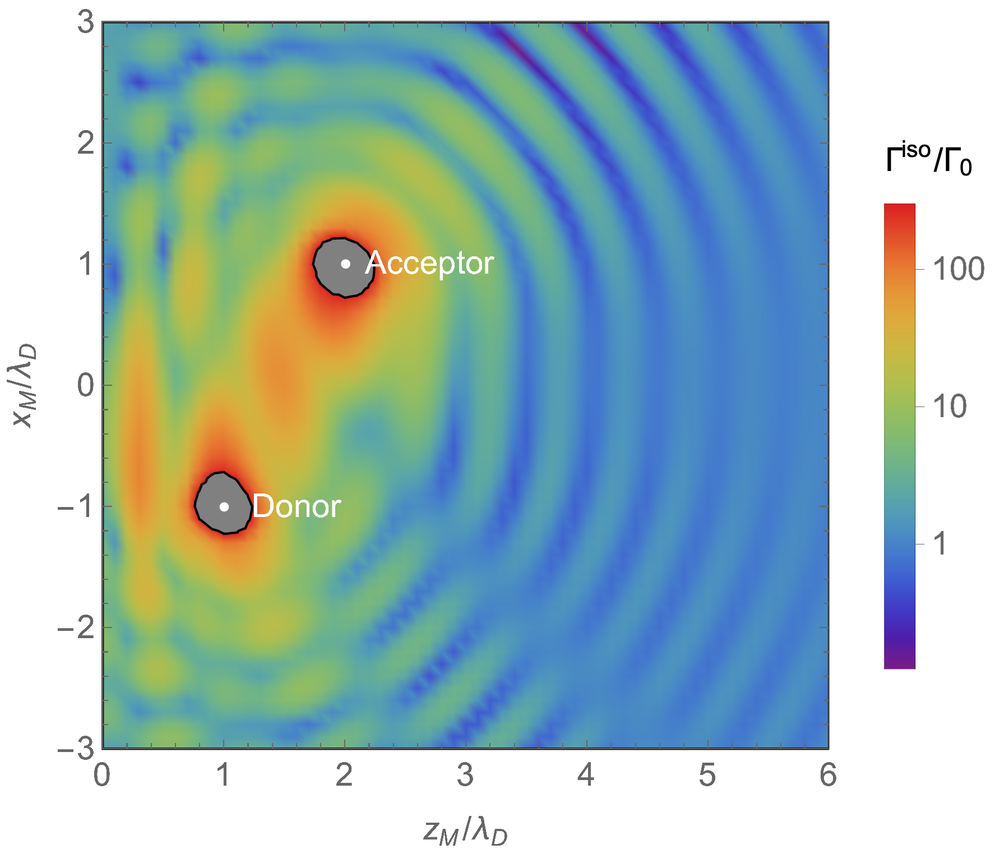}
  \caption{Rate of energy transfer for a donor and acceptor near a half-space with reflection coefficient $r_p=1$. The donor is fixed at position $\{x_\mathrm{D},z_\mathrm{D}\}/\lambda_\text{D} =\{-1,1\}$, the acceptor at position $\{x_\mathrm{A},z_\mathrm{A}\}/\lambda_\text{D} =\{1,2\}$, while the mediator is free to move in the $x$-$z$ pane. The other parameters and normalisation are the same as in Fig.~\ref{fig:colinearplot1}. The grey regions around donor and acceptor indicate where the rate enhancement goes off the colour scale.}
  \label{fig:contourplotrp=1}
\end{figure}

This points towards a highly non-trivial dependence of the donor-acceptor transfer rate when accompanied by a mediator and a nearby surface. To investigate this (and to go beyond the colinear case) we use the full form of the Green's tensor for an environment containing a half-space (given in \cite{buhmann_dispersion_2013}), of which the retarded \eqref{eq:G1R} and non-retarded \eqref{eq:G1NR} forms are  limits. A contour plot showing the rate for different positions of the mediator in the $x$- and $z$-axes is shown in Fig.~\ref{fig:contourplotrp=1}. The plot demonstrates the intricate dependence of the mediator's position on the rate of energy transfer between the donor and acceptor even in the presence of a relatively simple environment, producing both enhancement and suppression in different regions.

\section{Conclusions}
Here we have given a formula, \eqref{eq:rate}, which can be used to find the rate of mediated energy transfer in any external environment. This was derived using extended canonical perturbation theory beyond second order. It would have been possible to calculate this using standard perturbation theory and considering all time orderings, but this would have been an extremely complex and unwieldy process. We then applied this formula to a simple system, namely three bodies near an external semi-infinite half-space, but it could be applied to any environment for which the Green's tensor is known either analytically or numerically. 

The arrangement of a dimer made up of a donor and acceptor trapped near a surface controlled by an external agent is a situation of experimental interest (for example in \cite{tomasi_coherent_2019}), and we have shown how the presence and position of a third molecule influences the rate of energy transfer. Furthermore, long-range transfer in photosynthetic complexes may rely on the type of mediated RET discussed here. The formula presented here is also a minimal model of RET in a more complex environment and could be used as a starting point for such investigations.

The work presented here could also indicate a potential way to observe retardation in RET. Ordinarily the donor-acceptor rate at retarded distances is extremely small compared to the corresponding (observable) rates at smaller distances   \cite{andrews_intermolecular_1992}. However, adding a distant mediator to a non-retarded, surface enhanced reaction could be a way of observing the role of retardation in RET, without the complication of such low rates.

It is interesting to note that the form of the rate equation found, \eqref{eq:rate}, is exactly as one would anticipate from intuition about dipole moments and the Green's tensor. As indicated in Casimir and Polder's 1948 paper on interatomic potentials \cite{casimir_influence_1948}, this could likewise point towards a simpler way to obtain fully quantum formulae of this nature. This would be the start of a powerful method to carry out more complex many-body calculations.

\appendix 
\section{Consistency check with mirror-modified two-body rates} \label{ConsistencyAppendix}

In Eq.~(20) of Ref.~\cite{weeraddanaControllingResonanceEnergy2017}, a normal-mode QED-based formula is given for the mirror-modified two-body rate $\Gamma_\text{trans}(QD)$ for oriented (non-random) quantum dots modelled as dipoles. These are taken to be at a non-retarded distance from each other but arbitrary distance from a perfectly reflecting mirror. Similarly, the dipole moments in Ref.~\cite{weeraddanaControllingResonanceEnergy2017} are taken to be aligned with each other and with the surface of the mirror. Therefore, the (fully) non-retarded limit of Eq.~(20) in Ref.~\cite{weeraddanaControllingResonanceEnergy2017} should agree with with the perfect-reflector limit of the following special case of Eq.~\eqref{eq:rate};
\begin{align}
\Gamma_{xx} &=\frac{2\pi \mu_0^2 \omega_{\mathrm{D}}^4 }{\hbar}
|{d}^{\mathrm{A}*}_x  {G}^\text{NR}_{xx}(\bm{r}_{\mathrm{A}},\bm{r}_{\mathrm{D}},\omega_{\mathrm{D}})  {d}_x^{\mathrm{D}}|^2,
\end{align}
with ${G}^\text{NR}_{xx}$ being the $xx$ component of the sum of $\bm{G}_\mathrm{NR}^{(0)}$ given by Eq.~\eqref{eq:G0NR}, and $\bm{G}_\mathrm{NR}^{(1)}$ given by Eq.~\eqref{eq:G1NR}. Substituting these in and carrying out the algebra, we find;
\begin{align}\label{GammaXX}
\Gamma_{xx} =& \frac{|d_x^\mathrm{A}|^2|d_x^\mathrm{D}|^2 }{8 \pi  \hbar \epsilon_0^2} \Bigg[\frac{1}{\left(z_{\text{A}}-z_{\text{D}}\right){}^6}\notag \\
&-\frac{2 r_{\text{NR}}}{\left(z_{\text{A}}-z_{\text{D}}\right){}^3 \left(z_{\text{A}}+z_{\text{D}}\right){}^3}+\frac{r^{2 }_\text{NR}}{\left(z_{\text{A}}+z_{\text{D}}\right){}^6}\Bigg].
\end{align}
Translating the notation of Eq.~(20) in Ref.~\cite{weeraddanaControllingResonanceEnergy2017} to that used here (namely $R \to z_\mathrm{A}-z_\mathrm{D}$) gives
\begin{align}
&\Gamma_\text{trans}(QD)=\frac{|d_x^\mathrm{A}|^2|d_x^\mathrm{D}|^2 }{8\pi \hbar \epsilon_0^2} \Bigg[\frac{1}{\left(z_{\text{A}}-z_{\text{D}}\right){}^6}\notag \\
&-\frac{2 \cos \left(2 k z_{\text{D}}\right)}{\left(z_{\text{A}}-z_{\text{D}}\right){}^3 \left(z_{\text{A}}+z_{\text{D}}\right){}^3}+\frac{1}{\left(z_{\text{A}}+z_{\text{D}}\right){}^6}\Bigg],
\end{align}
the $k z_\mathrm{D}\to 0$ (non-retarded) limit of which exactly reproduces  the perfect reflector ($r_{\text{NR}} \to 1$) limit of Eq.~\eqref{GammaXX} above. 

\begin{acknowledgments}
The authors acknowledge financial support from the UK Research and Innovation Council through grants EPSRC/DTP 2020/21/EP/T517896/1 (M.C.W) and EP/V048449/1 (R.B.)
\end{acknowledgments}

\end{document}